\documentstyle[aaspp4,flushrt,psfig]{article}

\newcommand{\Mp}{{\cal M}_p}
\newcommand{\Mc}{{\cal M}_c}
\newcommand{\Mb}{{\cal M}_\bullet}

\newcommand{\msun}{M_\odot}
\newcommand{\tsun}{T_\odot}
\newcommand{\bm}[1]{{\mbox{\boldmath$#1$}}}

\lefthead{N. Wex \& S. M. Kopeikin}
\righthead{Frame dragging and other precessional effects \dots}

\begin{document}

\title{Frame dragging and other precessional effects\\ 
in black hole-pulsar binaries}

\author{N. Wex \altaffilmark{1,2}}

\and

\author{S. M. Kopeikin \altaffilmark{3,4}}

\altaffiltext{1}{Joseph Henry Laboratories and Physics Department, 
                 Princeton University, Princeton, New Jersey 08544, USA}
\altaffiltext{2}{MPI f\"ur Radioastronomie, Auf dem H\"ugel 69, 
                 53121 Bonn, Germany}
\altaffiltext{3}{Theoretisch Physikalisches Institut der 
                 Friedrich-Schiller-Universit\"at Jena,
                 Max-Wien-Platz 1, 07743 Jena, Germany}
\altaffiltext{4}{On leave from: PRAO, ASC FIAN, Leninskii Prospect 53, Moscow
117924, Russia}                  

\begin{abstract}

For radio pulsars in orbit with a compact companion, pulsar timing
observations have proved to be a powerful tool for identifying the physical
nature of the companion. Unfortunately, perhaps the most intriguing system
where such a tool could be used, a pulsar in orbit with a black hole, has yet
to be discovered.

In this paper we give a detailed investigation of what one can learn about the
black hole companion via timing observations of the pulsar.  We present an
analytic calculation for the propagation delay caused by the frame-dragging
effect and show that it has the same functional behavior as the modulation of
the observed rotational phase of the pulsar caused by the deflection of the
radio signals in the gravitational field of the companion (bending delay).
Thus, contrary to statements of other authors, the frame-dragging delay is
unlikely to be separately measurable in pulsar binaries where the companion is
a stellar mass black hole.  We demonstrate, however, that the precession of
the binary orbit caused by the relativistic spin-orbit coupling can lead to
observable effects which can be used to set a lower limit to the black-hole
spin, or possibly allow the determination of its magnitude and orientation.
We give parameter estimates for two possible systems, a $10 M_\odot$ black
hole in orbit either with a young ($\sim 0.1$ s) pulsar or with a millisecond
pulsar.  Finally, we discuss the measurability of the quadrupole moment of the
rotating black hole companion which would test the presence of a Kerr black
hole.  As an interesting side result of our calculations, we can give a
further argument why the companion of PSR J0045--7319 cannot be a Kerr black
hole.

\end{abstract}


\section{Introduction}

So far the best arguments for the existence of stellar mass black holes (BHs)
are based on dynamical mass estimates in X-ray binaries.  The measurement of
absorption-line velocities of the secondary star allows us to determine a lower
limit to the mass of the compact companion.  If the mass of the companion
exceeds the calculated maximum mass of a neutron star ($\sim 3 M_\odot$) we
call it a BH candidate (see \cite{Wij96} for a list of BH candidates).  But to
argue convincingly that the companion is indeed a BH one must rule out
possible alternatives to stellar mass BHs.  In case of cold, non-rotating
neutron stars (NSs) the permissible maximum mass depends on a
rather complicated equation of state (EOS) for condensed matter at and above
nuclear density.  Since our knowledge of the EOS for matter above nuclear
density is yet imperfect the maximum masses of NSs range widely, from $1.5
M_\odot$ to about $2.5 M_\odot$ (\cite{FIP89}; \cite{KEH89}; \cite{Lat&al90};
\cite{WG92}). This is well below the dynamical mass estimations for many of
the suspected BH binaries.  However, if we are very conservative and assuming
complete ignorance about the EOS above a density of $10^{14}$ g cm$^{-3}$,
dropping the causality condition (i.e.\ allowing a dispersive medium), and
allowing rotation, then the maximum mass goes up to about $14 M_\odot$
(\cite{FI87}). A compact object with a maximum mass of $14 M_\odot$ can explain
all the dynamical mass estimations in BH candidates.  There are other ways of
exceeding the dynamical mass limits, such as rather exotic compact stars like
Q-stars (\cite{BLS90}; \cite{MSN98}) or boson stars made from a strongly
self-interacting scalar field (\cite{CSW86}), the latter having in principle
no maximum-mass limit. Finally, one might even think of abandoning general
relativity as the correct theory for the description of the strong
gravitational fields inside a neutron star (although pulsar timing tests of
the strong equivalence principle support the validity of general relativity in
these strong field regime (\cite{DS91}; \cite{Wex97}).

Recent advances in the theory of accretion physics (e.g., the
advection-dominated accretion flow model) lead to the conclusion that the
matter accreting onto the compact companion in some of the BH candidates does
not hit a hard surface. This may indirectly indicate the presence of an event
horizon through which matter and radiation can fall in but from which nothing
escapes, a fundamental property of BHs (\cite{NGM97}). At present the concept
of an event horizon in these BH candidates has rather the character of a
``most natural explanation'' than of an inevitable necessity.  Certainly the
absence of a hard surface does not prove the presence of a BH as predicted by
Einstein's theory of gravity, i.e.\ a Kerr BH.

Astrophysical BHs are expected to rotate.  The spin of the BH gives rise to a
so called gravitomagnetic field (\cite{TH85}) which causes the relativistic
dragging of inertial frames in the vicinity of the BH. The dragging of
inertial frames has two important consequences for an accretion disc around
the BH. First, the radius of the last (marginally) stable orbit of the
accretion disk is a function of the BH spin (\cite{BPT72}). Secondly, any
deviation of particle motion from the equatorial plane will cause a precession
of the particle orbit (\cite{LT18}; \cite{BP75}; \cite{SH87}). In principle,
these effects allow the determination of the spin of the accreting BH using
observations of low-energy X-ray spectra (\cite{ZCC97}) or high frequency
quasi-periodic oscillations (\cite{CZC98}).  While it is still unclear whether
the observed quasi-periodic oscillations are a result of the Lense-Thirring
precession of inner parts of a warped accretion disk, spectral analysis seems
to show the presence of the gravitomagnetic field of the BH (\cite{ML98};
\cite{McCli98}). Nevertheless, the arguments are still require certain
plausible assumptions (e.g., the accretion disk reaches all the way down to
the last stable orbit, etc.) and the precision with which the spin is
determined is rather poor, even in systems where all the input parameters are
comparably well known.

The discovery (\cite{HT75}) and continuous observation of binary pulsars
(i.e.\ pulsars in orbit with a gravitationally bound companion) has opened an
entirely new testing ground for gravity theories.  So far binary pulsars
provide the only laboratories where one can probe the gravitational radiation
properties and strong field aspects of relativistic gravity (\cite{TW89};
\cite{DT92}; \cite{DE-F96a}; \cite{DE-F96b}). A useful review on pulsar timing 
and relativistic gravity can be found in \cite{Tay93}, where the reader can
find the basic ideas of this subject. 

\begin{figure}
\centerline{\psfig{figure=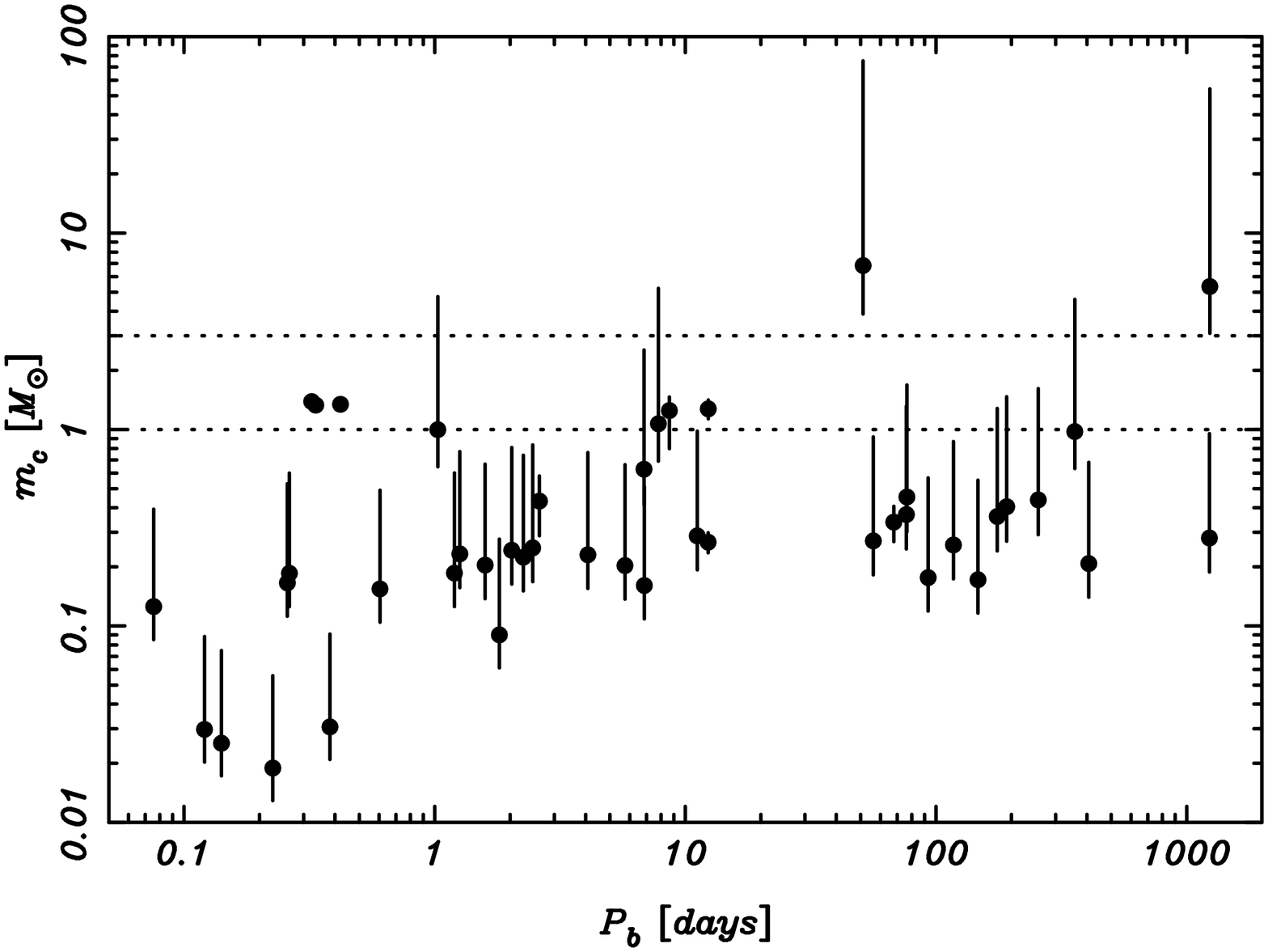,angle=-90,width=7.2in}}
\caption{Mass ranges of companion stars in pulsar binaries as estimated from
timing observations. The solid circles give the most likely value for the
companion mass while the error bars indicate a 95\% confidence level. The
lower horizontal dashed line indicates the Chandrasekhar mass limit, the upper
one indicates the 3 $M_\odot$ upper mass limit for neutron stars. There are
only five binary pulsars where the plotted mass range of the companion exceeds
3 $M_\odot$. The first two binary pulsars from the left (B0655$+$64,
J1022$+$1001) have optically detected white dwarf companions (Kulkarni 1986;
Lundgren, Foster, \& Camilo 1996). The third binary pulsar (J0045$-$7319) has
a 9 $M_\odot$ B-star companion (Bell {\it et al.}\ 1995), and the fifth binary
pulsar (B1259--63) is in orbit with a 10 $M_\odot$ Be-star (Johnston {\it et
al.}\ 1994). Only the nature of number four (B1820$-$11) is unclear.  Mass
functions and masses were taken from (Deich \& Kulkarni 1996, Nice, Sayer \&
Taylor 1996, Stairs {\it et al.}\ 1998, Kaspi, Taylor \& Ryba 1994, Taylor
{\it et al.}\ 1993, 1995). We further assumed that the mass of the pulsar is
above 1.3 $M_\odot$}
\label{f1}
\end{figure}

A pulsar orbiting a BH in a close orbit would certainly be a high precision
laboratory for BH physics (\cite{PT79}). After a short observation period it
would be possible to measure the relativistic advance of periastron,
$\dot\omega$, and thus obtain the total mass of the binary system to high
precision. At this stage the mass of the BH would be fairly well constrained
since the mass of the pulsar is unlikely to exceed $2\msun$. The measurement
of a second post-Keplerian (PK) parameter would then yield the mass of
the pulsar, $m_p$ ($\equiv\Mp\msun$), and the mass of the companion, $m_c$
($\equiv\Mc\msun$). It was even suggested (\cite{NP91}, \cite{LW97}) that a
pulsar in a close, nearly edge-on orbit, with a spinning BH might be used to
probe the rotation of the BH via the timing delay produced by the dragging of
inertial frames caused by the spin of the BH (frame-dragging propagation
effect). Unfortunately so far none of the binary pulsars seems to orbit a
stellar mass BH of typically $\ga 10 M_\odot$ (see Fig.~\ref{f1}).

However, it is possible that pulsars in close orbit with BHs have already been
missed in previous pulsar surveys due to the Doppler smearing of the signal
during the integration. Whilst it is true that this effect can be largely
removed by the application of an ``acceleration code'' (\cite{JK91}), only a
number of globular-cluster searches have utilized this technique due to
previous limitations in computing power (\cite{And92}). The continual
improvement and availability of more powerful computers, however, means that
many more surveys are now able to incorporate acceleration searches 
(\cite{Lor98}).

In this paper we discuss what one can learn about a stellar mass BH by timing
its pulsar companion. We focus, in particular, on the observational
determination of the BH's rotation (angular velocity and orientation) and
assume throughout the paper that in addition to the five Keplerian parameters,
at least two PK parameters are measured with high precision, so that we know
$\Mp$, $\Mc$, and therefore $\sin i$, i.e.\ the angle of orbital inclination,
$i$, up to the ambiguity $i\rightarrow\pi-i$. We begin with a brief
introduction to the definition of the spin and quadrupole moments of a
rotating body in general relativity and relations between them (\S 2).  In \S
3 we give an analytic treatment of the frame-dragging propagation effect and
show that it is practically impossible to observe this effect in stellar mass
BH-pulsar binaries.  In \S 4 we concentrate on the orbital dynamics of a
pulsar orbiting a rotating BH and give a detailed investigation of the
precession of the pulsar orbit caused by the spin (relativistic spin-orbit
coupling) and quadrupole moment (classical spin-orbit coupling) of the BH. In
\S 5 we calculate the secular changes in observable quantities caused by the
relativistic spin-orbit coupling. We show that, in principle, one can extract
the spin of the BH from nonlinear-in-time changes of the observables which can
be approximated by polynomials in time with sufficient accuracy.  We discuss
additional effects that can cause secular changes in the observed parameters
reducing the measurement accuracy for the BH spin.  In \S 6 we outline the
prospects of BH spin determination once a BH-pulsar binary is discovered.  In
\S 7 we present a method of extracting the quadrupole moment of the rotating
BH from the orbital dynamics of the pulsar and evaluate its actual
measurability. We summarize in \S 8.


\section{Spin and quadrupole moment of a compact body}

The external metric of a stationary, axially symmetric body can be
written in the standard form (\cite{BW71})
\begin{equation}
   ds^2 = -e^{2\nu}\;c^2dt^2+R^2\sin^2\theta B^2e^{-2\nu}(d\phi-\omega\;dt)^2 
          +e^{2\alpha} (dR^2 + R^2\; d\theta^2) \;, 
\end{equation}
where $c$ is the speed of light; $R$, $\phi$, $\theta$ are ordinary spherical
coordinates, and the potentials $\nu$, $B$, $\omega$, and $\alpha$ are
functions of $R$ and $\theta$. Butterworth and Ipser (1976) have calculated
the asymptotic behavior of these potentials for large $R$. In particular, the
asymptotic behavior of the potentials $\nu$ and $\omega$ is (using $c\equiv
G\equiv 1$)
\begin{eqnarray}
   \nu    &=& -\frac{M}{R}-\frac{1}{12}\left(\frac{M}{R}\right)^3
              - q\left(\frac{M}{R}\right)^3 P_2(\cos\theta) 
              + O\left(\frac{1}{R^4}\right) \;,\label{asympt1}\\ 
   \omega &=& 2\chi\left(\frac{M}{R}\right)^3 
              + O\left(\frac{1}{R^4}\right) \;,\label{asympt2} 
\end{eqnarray}
(\cite{BI76}, \cite{LP97}); $M$ is the body's mass and $\chi$ is a
dimensionless measure of the body's angular momentum (spin), $S$:
\begin{equation}\label{spinbh}
   \chi \equiv \frac{c}{G}\; \frac{S}{M^2} \;.
\end{equation}
The dimensionless parameter $q$ in equation (\ref{asympt1}) is related to the
quadrupole moment, $Q$, of the rotating body:
\begin{equation}\label{quadrubh}
   q = \frac{c^4}{G^2}\; \frac{Q}{M^3} \;.
\end{equation}
In the Newtonian limit one has
\begin{equation}\label{newton}
   Q = \int \varrho(R',\theta')\, R'^2\, P_2(\cos\theta')\, dV' \;,
\end{equation}
where $\varrho$ is the mass density of matter inside the star. The Newtonian
theory of self-gravitating, rotating bodies predicts a certain relationship
between the angular velocity of the body's rotation and its quadrupole moment
due to the oblateness in the mass distribution (\cite{Ch69}). For the same
reason, one now may expect a complicated relation between the rotational
parameter of a star, $\chi$, and its quadrupole moment, $q$, in relativistic
gravity theories. For a rotating Kerr BH in general relativity $\chi\le1$ for
if one had $\chi>1$ it would give an unacceptable naked singularity (see
\cite{HE73}).  Furthermore, for a Kerr BH the quadrupole parameter, $q$, is
uniquely determined by the rotational parameter $\chi$ (\cite{Tho80},
\cite{TPM86}):
\begin{equation}\label{nohair}
   q = -\chi^2. 
\end{equation}

Similarly for solid extended bodies, Laarakkers and Poisson (1997) have
computed the quadrupole moment of rotating neutron stars in the interval
between 1.0 and 1.8 solar masses using four different EOS. They have found the
maximum values for $\chi$ lying in between 0.62 (softest EOS) and 0.73
(stiffest EOS). For the quadrupole moment of the mass distribution, $q$, they
have derived a dependence on $\chi$ which is well reproduced by the relation
\begin{equation}\label{bhqchi}
   q \simeq -C\chi^2 \;,
\end{equation}
where the constant $C$ takes a value between 2.0 and 12.1 depending on the
mass of the NS and the EOS.

For spinning boson stars with large self-interaction the maximum value for the
parameter $\chi$ seems to be between 3 and 4, and the relation between $\chi$
and $q$ is rather complicated once $\chi$ exceeds 0.2 (\cite{Rya97}). The 
following inequality, however, always holds:
\begin{equation}\label{bsqchi}
   q \la - 10 \chi^2.
\end{equation}
Comparing Eqs. (\ref{nohair}) -- (\ref{bsqchi}) one sees that simultaneous
measurements of the mass, $M$, spin parameter, $\chi$, and the quadrupole
moment, $q$, of the pulsar's compact companion can lead to a unequivocal
identification of its physical nature.\footnote{We are not aware of any
calculations done for Q-stars.} Such measurements seem to be a straightforward
way to observationally verify the existence of a rotating Kerr BH.


\section{The frame-dragging propagation effect in highly inclined binary
         systems}

In Newtonian gravity the rotation of a star contributes only indirectly to its
gravitational field through the rotationally induced oblateness of the mass
distribution. The centrifugal flattening of the surfaces of equal mass density
caused by the rotation of the star gives rise to a gravitational quadrupole
field. In general relativity, however, not only mass itself contributes to the
gravitational field. The rotation of a star gives rise to a mass current which
contributes in a specific way to its gravitational field. The rotation is the
source of the {\it gravitomagnetic field} of the rotating body (\cite{TPM86}).
This gravitomagnetic field influences the motion of test particles and the
propagation of light in the vicinity of the rotating body (see, for instance
\cite{CW95}, \cite{M97} and references therein).  It is well-known that BHs,
although purely geometrical in their nature (i.e.\ a solution of Einstein's
vacuum-field equations), are physical objects and as such also can carry
angular momentum which gives rise to a gravitomagnetic field. In fact, the
strength and configuration of the gravitomagnetic field of the BH is used to
define its rotation. The rotation of the BH will influence the propagation of
photons in a specific way (\cite{Bar73}). For pulsar-timing experiments its
influence on the propagation time of radio signals are of particular
interest. We call this contribution to the propagation time of the pulsar
signals the {\it frame-dragging propagation effect}.

Recently, Laguna and Wolszczan (1997) have suggested on the basis of numerical
ray-tracing calculations that pulsar-timing experiments could measure the
gravitomagnetic field of a rotating BH companion by its influence on the
propagation of the pulsar signals emitted from different points of the
pulsar's orbit while the pulsar is passing through the point of the superior
conjunction with the companion.  In this section we give a more detailed
investigation of this frame-dragging propagation effect based on analytic
calculations and show the problems preventing its direct
measurement.

For a discussion of the effect in question it is convenient to use the metric
of Kerr spacetime, written in terms of Boyer-Lindquist time $t$ and
asymptotically Cartesian spatial coordinates $X^j$. One has (\cite{TPM86}):
\begin{equation}
   ds^2=-\alpha^2c^2dt^2+g_{jk}(dX^j+\beta^jdt)(dX^k+\beta^kdt) \;,
\end{equation}
where at large distances from the BH one finds for the lapse, $\alpha$, and
shift, $\beta^i$, functions the following expressions (\cite{TPM86}):
\begin{equation}
   \alpha^2 = 1+{\cal O}\left(\frac{1}{R}\right) \;, \qquad
   \bm{\beta} = -\frac{2G}{c^2}\;\frac{{\bf S}\times{\bf X}}{R^3} 
                +{\cal O}\left(\frac{1}{R^3}\right)\;, \qquad 
   g_{jk}=\delta_{jk}+{\cal O}\left(\frac{1}{R}\right) \;,
\end{equation}
where $R=|{\bf X}|$ is the distance from the BH to the current point in space.
The shift function $\bm{\beta}$ represents the vector potential of the
gravitomagnetic vector field which drags local inertial frames into rotation
(frame dragging effect). Since for photons $ds^2=0$ the contribution of the
frame dragging effect to the propagation time of a photon from the pulsar to
the observer is
\begin{equation}\label{dtFD}
   dt_{FD} \simeq \frac{\bm{\beta}\cdot d{\bf X}}{c^2} = -\frac{2G}{c^4}\;
        \frac{({\bf S}\times{\bf X})\cdot d{\bf X}}{R^3} \;.
\end{equation}
After substituting into the right-hand side of equation (\ref{dtFD}) the
unperturbed trajectory of the photon, which is a straight line, the equation
is easily integrated in a coordinate system where ${\bf X}=(x,0,b)$. The
constant $b$ is the minimum distance of the unperturbed trajectory of the
photon to the BH.\footnote{Details about the propagation of light rays in
the stationary field of relativistic gravitational multipoles can be found in
(\cite{Kop97a})} Consequently, the timing delay caused by the frame-dragging
effect assumes the form
\begin{equation}\label{infr}
    \Delta_{FD}=-\frac{2G}{c^4}\;S_yb\int_{x_0}^{\infty}
                 \frac{dx}{(x^2+b^2)^{3/2}}
       =-\frac{2G}{c^4}\;\frac{S_y}{b}\left(1-\frac{x_0}
       {r_0}\right) \;,
\end{equation}
where $r_0=|{\bf X}_0|$ is the distance between the BH and the point of
emission of the photon which is located at the pulsar's orbital position.
We have assumed that the observer is far away from the
rotating BH; for this reason, the upper limit of the integral in (\ref{infr})
was taken to be infinity. Indeed, from equation (\ref{dtFD}) it is
already clear that most of the contribution to this {\it frame-dragging (FD)
propagation effect} comes from a region around the BH with a few impact
parameters, $b$, in radius.

\begin{figure}
\centerline{\psfig{figure=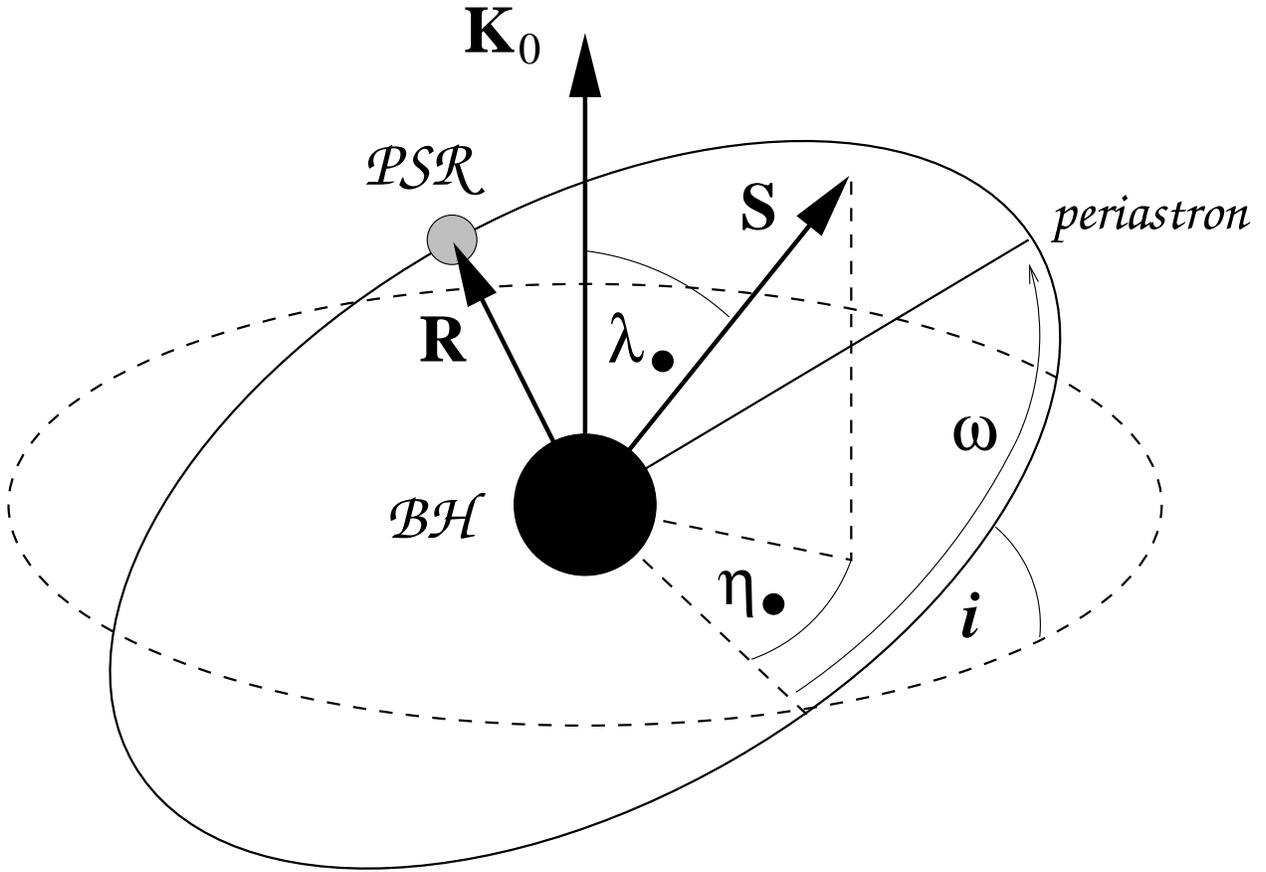,width=6.5in}}
\caption{Orientation of the BH-PSR system in the sky. The vector ${\bf S}$ is
the spin of the BH, ${\bf R}$ is the vector pointing from the BH to the
pulsar. The angle $\lambda_\bullet$ is that between the BH spin and the
direction of the line-of-sight given by the unit vector ${\bf K}_0$. The angle
$\eta_\bullet$ is that between the projection of the BH spin on to the plane
of the sky and the ascending node of the binary orbit.}
\label{f2}
\end{figure}

Let ${\bf R}$ denote the vector pointing from the BH to the pulsar and ${\bf
K}_0$ denote the unit vector pointing from the observer to the pulsar (see
Fig.~\ref{f2}).  Then, ${R}\equiv|r_0|$, and one has (cf. eq. (43) for $l=1$
in \cite{Kop97a})
\begin{equation}
   \Delta_{FD} = -\frac{2G}{c^4}\;\frac{{\bf S}\cdot({\bf K}_0\times{\bf R})}
         {R(R-{\bf K}_0\cdot{\bf R})} \;.
\end{equation}
Using the angles as defined in Fig.~\ref{f2} we find for the FD
propagation effect in a binary pulsar system
\begin{equation}\label{FDPE}
   \Delta _{FD}=\Lambda^{-1}\biggl\{{\cal A}_{FD}\cos
   i\sin\left[\omega+A_e(u)\right]
                +{\cal B}_{FD}\cos\left[\omega+A_e(u)\right]\biggr\}\;,
\end{equation}
where the function $\Lambda$ is defined by
\begin{equation}
   \Lambda=1-e\cos u-s[\sin\omega(\cos u-e)+(1-e^2)^{1/2}\cos\omega\sin u]\;,
\end{equation}
the eccentric anomaly angle is
\begin{equation}
   A_e(u) = 2\arctan\left[\left(\frac{1+e}{1-e}\right)^{1/2}
         \tan\left(\frac{u}{2}\right)\right] \;,
\end{equation}
and 
\begin{eqnarray}
   {\cal A}_{FD}&=& +4\;\tsun^{5/3}\left(\frac{2\pi }{P_b}\right)^{2/3}
            \frac{{\cal M}_\bullet^2}{({\cal M}_p+{\cal M}_\bullet)^{1/3}}\;
            \chi\sin\lambda_\bullet\cos\eta_\bullet\;,
            \label{FDPEC1}\\
   {\cal B}_{FD}&=& -4\;\tsun^{5/3}\left(\frac{2\pi}{P_b}\right)^{2/3}
            \frac{{\cal M}_\bullet^2}{({\cal M}_p+{\cal M}_\bullet)^{1/3}}\;
            \chi\sin\lambda_\bullet\sin\eta_\bullet\;.
            \label{FDPEC2}
\end{eqnarray}
When ${\cal M}_p\ll{\cal M}_\bullet$ we find the following numerical
estimate for the constant factor
\begin{equation}
   \tsun^{5/3}\left(\frac{2\pi}{P_b}\right)^{2/3}
   \frac{{\cal M}_\bullet^2}{({\cal M}_p+{\cal M}_\bullet)^{1/3}} \approx
   (0.0001\;\mu{\rm s})\left(\frac{P_b}{1\;{\rm day}}\right)^{-2/3}
   \left(\frac{{\cal M}_\bullet}{10}\right)^{5/3}.
\end{equation}

To illustrate the expected strength of the FD propagation effect we assume a
pulsar in a circular orbit where $\eta_\bullet=0$ and the inclination angle
$i$ is close to $90^\circ$, i.e.\ $e=0$, $\omega=0$, $A_e=u$, $s=\sin i\simeq
1$, and $\cos i\ll 1$. Then ${\cal A}_{FD}\neq 0$, ${\cal B}_{FD}= 0$, and
\begin{equation}
   \frac{\sin(\omega+A_e)}{\Lambda}=\frac{\sin u}{1-s \sin u}\;.
\end{equation}
This function has maximum at $u=\pi/2$ and minimum at $u=3\pi/2$. 
Thus, for the maximal magnitude of the FD effect we have: 
\begin{equation}\label{estFD}
{\rm max}\left(\Delta _{FD}\right)\approx
   \frac{0.0008\;\mu{\rm s}}{|\cos i|}
   \left(\frac{P_b}{1\;{\rm day}}\right)^{-2/3}
   \left(\frac{{\cal M}_\bullet}{10}\right)^{5/3}\chi\sin\lambda_\bullet\;.
\end{equation}
Hence, for BH binary pulsars with an orbital inclination $i$ very close to
$90\arcdeg$, the FD propagation effect will have a measurable influence on the
TOAs, provided that the pulsar is a millisecond pulsar where one can expect
(with present-day technology) a timing accuracy of better than one microsecond.
This fact was already pointed out by Laguna \& Wolszczan (1997) using
numerical methods to study the propagation of the pulsar signals in the
spacetime of the Kerr-BH companion. In particular, for a binary pulsar with an
orbital period $P_b$ about 1 day and an extreme $10M_\odot$ Kerr BH companion
with $\chi=1$ the strength of the FD propagation effect will be one
microsecond if $\cos i\leq 0.001$ or $|i-90\arcdeg| \la 0.05\arcdeg$.

A measurement of at least one of the parameters ${\cal A}_{FD}$ or ${\cal
B}_{FD}$ will give values for either $\chi \sin\lambda_\bullet
\cos\eta_\bullet$ or $\chi \sin\lambda_\bullet \sin\eta_\bullet$. Since we
have a priori no idea about the orientation of the BH spin it is not possible
to extract $\chi$ from these measurements. But we get a lower limit on
$|\chi|$, which has to be less than one, otherwise it is not a Kerr BH.  In
the lucky case of having good values for both parameters, ${\cal A}_{FD}$ and
${\cal B}_{FD}$, we can calculate $\eta_\bullet$ and consequently
$\chi\sin\lambda_\bullet$ which gives a definite lower limit for the $\chi$ of
the companion. An upper limit to $\chi$ can be set only with a certain
probability excluding values of $\lambda_\bullet$ being close to either
$0\arcdeg$ or $180\arcdeg$.

Unfortunately, the measurement of the FD propagation effect is complicated (and
may be impossible) by a competing effect which occurs during the superior
conjunction, the {\it bending delay} (\cite{DK95}) The delay originates
because of the modulation of the pulsar's rotational phase by the effect of
gravitational deflection of light in the field of the pulsar's companion. The
expression for the bending delay in the framework of general relativity is
given by the formula (\cite{DK95})
\begin{equation}\label{BD}
\Delta _B=\Lambda^{-1}[{\cal A}_B\cos i\sin (\omega+A_e)
                +{\cal B}_B\cos(\omega +A_e)] \;,
\end{equation}
where 
\begin{eqnarray}
{\cal A}_B &=& +\frac{\tsun^{2/3}}{\pi\nu_p}\;
             \left(\frac{2\pi}{P_b}\right)^{2/3}
             \frac{{\cal M}_\bullet}{({\cal M}_p+{\cal M}_\bullet)^{1/3}}\;
             \frac{\cos \eta_p}{\sin \lambda _p} \;, \label{BDC1}\\
{\cal B}_B &=& -\frac{\tsun^{2/3}}{\pi\nu _p}\;
             \left(\frac{2\pi}{P_b}\right)^{2/3}
             \frac{{\cal M}_\bullet}{({\cal M}_p+{\cal M}_\bullet)^{1/3}}\;
             \frac{\sin \eta _p}{\sin\lambda _p} \;. \label{BDC2}
\end{eqnarray}
Here, $\nu_p$ is the rotational frequency of the pulsar, and the angles
$\lambda_p, \eta_p$ have the same meaning as $\lambda_{\bullet},
\eta_{\bullet}$ but define now the orientation of the pulsar's angular
velocity vector in space. For $\Mp\ll\Mb$ we find
\begin{equation}
   \frac{\tsun^{2/3}}{\pi\nu_p}\left(\frac{2\pi}{P_b}\right)^{2/3}
   \frac{{\cal M}_\bullet}{({\cal M}_p+{\cal M}_\bullet)^{1/3}} \approx
   (0.0007\;\mu{\rm s})\left(\frac{P}{1\;{\rm ms}}\right)
   \left(\frac{P_b}{1\;{\rm day}}\right)^{-2/3}
   \left(\frac{{\cal M}_\bullet}{10}\right)^{2/3}\;,
\end{equation}
where $P$ denotes the rotational period of the pulsar.

The bending delay has the same functional dependence on the orbital motion as
the FD propagation effect and, for this reason, one can observe only a linear
combination of the FD effect and the bending delay.  Hence, in timing
observations the parameters of the FD effect are not separated from those of
the bending delay.  For pulsars with rotational periods close to 1 ms the
amplitudes of these two effects are comparable if ${\cal M}_\bullet\sim
20$. For smaller BH masses and/or longer pulsar periods the bending delay is
clearly the dominating effect. Only for BH companions with more than $\sim 100
M_\odot$ the FD propagation effect will dominate over the bending delay.

In principle, one may extract $\eta_p$ and $\lambda_p$ from additional pulse
structure analysis (\cite{DT92}). In this case the bending delay can be
predicted and subtracted from the total effect. In practice, pulse structure
observations and subsequent analysis are connected with rather large
uncertainties in the corresponding angles. Therefore, while using results of
such an analysis one can still hope to get a comparably good measurement for
$\lambda_p$, the angle $\eta_p$ affects the polarization pattern of the pulsar
only at the negligible small $x/P_b$ level (\cite{DT92}) and, thus, will
remain unobservable.


\section{Spin-orbit coupling and precession of the binary orbit}

Eighty years ago Lense \& Thirring (1918) pointed out that the gravitomagnetic
field of a central rotating body will cause a precession of the orbit of a
test particle. In the same way, the rotation of one or both components of a
binary system will cause a precession of the binary orbit (\cite{Br91}). In
this and the following sections we will show that the observation of such a
precession can lead to the direct determination of the spin of a BH companion.

The typical moment of inertia for a pulsar is of order $10^{45}\;{\rm
cm}^2{\rm g}$ (\cite{AB77}).  Thus, assuming the rotational period of the
pulsar to be just 1 ms we obtain a spin of order $6\times10^{48}\;{\rm
cm}^2{\rm g/s}$. Using equation (\ref{spinbh}) we find that the spin of a $10
M_\odot$ extreme Kerr BH ($\chi=1$) has a value more than 100 times
bigger. Since for even a very soft EOS the rotational period of a pulsar
should lie above $\sim$0.5 ms (\cite{WG92}), the spin of the BH will dominate
the orbital precession.  For this reason, we neglect the spin of the pulsar in
our subsequent calculations.  The secular precession of the orbit is then
given by two vector differential equations of first order (\cite{Br91})
\begin{equation}
   \dot{\bf L}=\bm{\Omega}_{\rm prec}\times{\bf L} \;, \qquad
   \dot{\bf A}=\bm{\Omega}_{\rm prec}\times{\bf A} \;,
\end{equation}
where ${\bf L}$ is the orbital angular momentum and ${\bf A}$ is the
Laplace-Runge-Lenz vector which points to the instantaneous position of
periastron of the precessing orbit. The angular velocity vector
$\bm{\Omega}_{\rm prec}$ is the sum of the well-known relativistic
(post-Newtonian) periastron advance, $\bm{\Omega}_{PN}$, the gravitomagnetic
Lense-Thirring precession, $\bm{\Omega}_S$, caused by the coupling of the
orbital angular momentum vector to the spin of the BH, and the classical
precession, $\bm{\Omega}_Q$, due to the Newtonian coupling of the orbital
angular momentum vector to the quadrupole moment of the rotating BH
\begin{equation}
   \bm{\Omega}_{\rm prec}=\bm{\Omega}_{PN}+\bm{\Omega}_S+\bm{\Omega}_Q \;,
\end{equation}
where
\begin{eqnarray}
   \bm{\Omega}_{PN} &=& \Omega^\star_{PN}\;\hat{\bf L}\;, \\
   \bm{\Omega}_S &=& \Omega^\star_S\;[3(\hat{\bf L}\cdot\hat{\bf S})
                      \hat{\bf L}-\hat{\bf S}]\;, \\
   \bm{\Omega}_Q &=& \Omega^\star_Q\;[\{5(\hat{\bf L}\cdot\hat{\bf S})^2-1\}
                     \hat{\bf L}-2(\hat{\bf L}\cdot\hat{\bf S})\hat{\bf S}]\;,
\end{eqnarray} 
and a hat on a vector indicates the unit vector in the same direction as
that of the vector itself. It is important to stress that $\Omega^\star_S$ and
$\Omega^\star_Q$ are not equal to the absolute values of the corresponding
vectors. (We have deliberately used the asterisk to avoid such a possible
confusion.) One finds (see \cite{BOC75}):
\begin{eqnarray}
   \Omega^\star_{PN} &=& \frac{\tsun^{2/3}}{1-e^2}\left(\frac{2\pi}{P_b}
        \right)^{5/3}3(\Mp+{\cal M}_\bullet)^{2/3}\;,\;\label{Omstar1}\\
   \Omega^\star_S &=& -\frac{\chi\;\tsun}{(1-e^2)^{3/2}}\;
        \left(\frac{2\pi}{P_b}\right)^2
        \frac{{\cal M}_\bullet(3{\cal M}_p+4{\cal M}_\bullet)}
        {2({\cal M}_p+{\cal M}_\bullet)}\;, \\
   \Omega^\star_Q &=& -\frac{q\;\tsun^{4/3}}{(1-e^2)^2}\;
        \left(\frac{2\pi}{P_b}\right)^{7/3}
        \frac{3{\cal M}_\bullet^2}{4({\cal M}_p+{\cal M}_\bullet)^{2/3}} \;.
        \label{Omstar3}
\end{eqnarray}
It is worth emphasizing that both vectors ${\bf L}$ and ${\bf S}$ precess
around the conserved total angular momentum of the binary system ${\bf
J}\equiv{\bf L}+{\bf S}$, while their absolute values, $L\equiv|{\bf L}|$ and
$S\equiv|{\bf S}|$, are conserved quantities (averaged over one orbital
revolution).

\begin{figure} 
\centerline{\psfig{figure=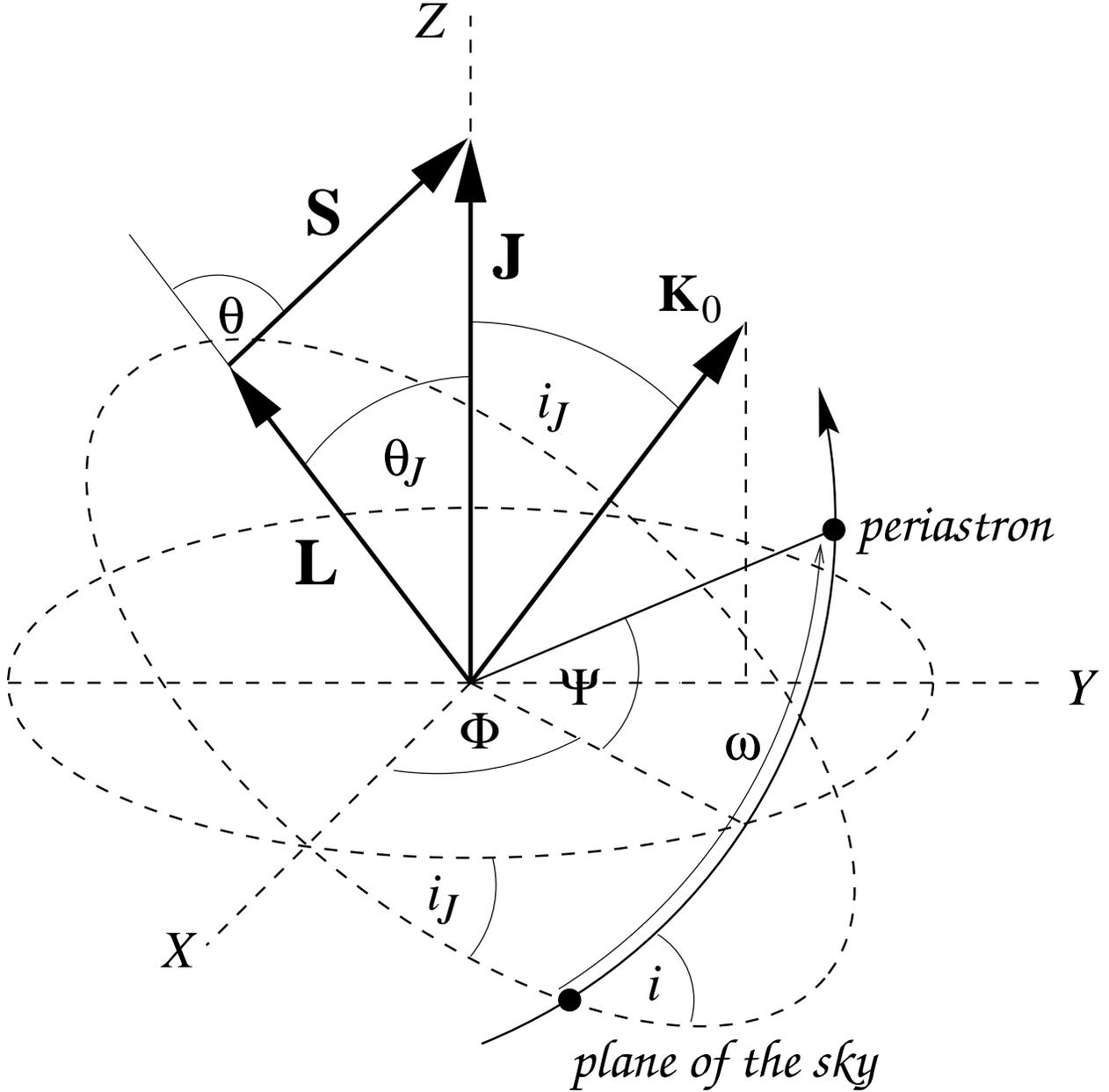,width=6.5in}}
\caption{Definition of angles in the total-angular-momentum reference frame.
The invariable ($X$-$Y$) plane is perpendicular to the total angular momentum
${\bf J}={\bf L}+{\bf S}$. The line-of-sight vector ${\bf K}_0$ is in the
$Y$-$Z$ plane. The vector ${\bf J}$ is a conserved quantity and, if averaged
over one full orbital period, the absolute values $|{\bf L}|$ and $|{\bf S}|$ 
are constant. Thus, the angles $i_J$, $\theta_J$, and $\theta$ are fixed. The
angles $\Phi$ and $\Psi$ change linearly with time.}
\label{f3}
\end{figure}

The precession of the orbital plane and the longitude of periastron are best
described by the angles $\Phi$ and $\Psi$ defined with respect to
the invariable plane, i.e.\ the plane perpendicular to ${\bf J}$ (see
Fig.~\ref{f3} for details). From the identities
\begin{eqnarray}
   &&\bm{\Omega}_{\rm prec}\times{\bf L}=\dot{\bf L}
    =\dot\Phi\hat{\bf J}\times{\bf L} \;,\\
   &&\bm{\Omega}_{\rm prec}\times{\bf A}=\dot{\bf A}
    =(\dot\Phi\hat{\bf J}+\dot\Psi\hat{\bf L})\times{\bf A}
\end{eqnarray}
one obtains the precession in terms of the angles $\Phi$ and $\Psi$:
\begin{equation}
   \dot\Phi = \dot\Phi_S+\dot\Phi_Q = {\rm const.} \;,
\end{equation}
where
\begin{eqnarray}
   \dot\Phi_S &=& -\Omega^\star_S\left(\frac{\sin\theta}{\sin\theta_J}
      \right) \;, \\
   \dot\Phi_Q &=& -\Omega^\star_Q\left(\frac{\sin2\theta}{\sin\theta_J}
      \right)\;,\label{phiq}
\end{eqnarray}
and
\begin{equation}
   \dot\Psi = \dot\Psi_{PN}+\dot\Psi_S+\dot\Psi_Q = {\rm const.} \;,
\end{equation}
where
\begin{eqnarray}
   \dot\Psi_{PN} &=& \Omega^\star_{PN}=\dot\omega_{GR0} \;, \\ 
   \dot\Psi_S &=& \Omega^\star_S(2\cos\theta+\sin\theta\cot\theta_J) \;,\\
   \dot\Psi_Q &=& \Omega^\star_Q\left(\case{1}{2}+\case{3}{2}\cos2\theta
                  +\sin2\theta\cot\theta_J\right) \;.\label{psiq}
\end{eqnarray}
In particular, eqs. (\ref{phiq}), (\ref{psiq}) can be found in Smarr \&
Blandford (1976) (see also \cite{LBK95}).

It is worth noting that the rate of the precession depends on two angles
$\theta$ and $\theta_J$. However, given the spin of the BH, $S$, one can
express $\theta_J$ as a function of $\theta$, since
$L\sin\theta_J=S\sin(\theta-\theta_J)$.  We find
\begin{eqnarray}\label{theta}
   \frac{1}{\sin\theta_J} &=& \frac{1}{\sin\theta}\;
   \left(1+\frac{L^2}{S^2}+2\,\frac{L}{S}\,\cos\theta\right)^{1/2} \;,
\end{eqnarray}
\begin{eqnarray}
   \cot\theta_J &=& \cot\theta+\frac{L}{S\sin\theta} \;,
\end{eqnarray}
where
\begin{equation}
   \frac{S}{L}=\frac{\chi\;\tsun^{1/3}}{(1-e^2)^{1/2}}\;
               \left(\frac{2\pi}{P_b}\right)^{1/3}\;
               \frac{\Mb(\Mp+\Mb)^{1/3}}{\Mp}\;.
\end{equation}
Hence, the rate of the orbital precession may be expressed as a function of
only one angle.

We conclude this section by giving numerical values for the various
precessional effects.  Assuming a pulsar mass of $1.4 M_\odot$ and a Kerr BH
mass ${\cal M}_\bullet \ga 10$ allows us to approximate equations
(\ref{Omstar1}) to (\ref{Omstar3}) by the expressions
\begin{eqnarray}
   \Omega^\star_{PN} &\approx&\frac{1}{1-e^2}\; 
    \left(\frac{P_b}{1\;{\rm day}}\right)^{-5/3}
    \left(\frac{{\cal M}_\bullet}{10}\right)^{2/3} \;{\rm deg/yr}\;,\\
   \Omega^\star_S &\approx&0.9\times10^{-3} \;\frac{\chi}{(1-e^2)^{3/2}}\;
    \left(\frac{P_b}{1\;{\rm day}}\right)^{-2}
    \left(\frac{{\cal M}_\bullet}{10}\right) \;{\rm deg/yr}\;,\\
   \Omega^\star_Q &\approx&0.5\times10^{-6} \;\frac{\chi^2}{(1-e^2)^2}\;
    \left(\frac{P_b}{1\;{\rm day}}\right)^{-7/3}
    \left(\frac{{\cal M}_\bullet}{10}\right)^{4/3} \;{\rm deg/yr}\;.
    \label{OMqn}
\end{eqnarray}
Now it is easy to see that the precession in $\Psi$ is clearly dominated by
$\Omega^\star_{PN}$ while that in $\Phi$ is dominated by the relativistic
spin-orbit coupling $\Omega^\star_S$.  The precession $\Omega^\star_Q$ caused
by the quadrupole moment of the BH is three orders of magnitude smaller than
the relativistic spin-orbit precession and, therefore, will be omitted from
the following discussion.


\section{Spin-orbit coupling and observable quantities}

The angles $\Phi$ and $\Psi$, which change linearly in time, are not directly
observable in analyzing pulse arrival times.  Instead, one can extract from
the timing observations the projected semi-major axis of the pulsar orbit,
$x=a_p\sin i/c$, and the longitude of periastron, $\omega$, which are
connected with $\Phi$ and $\Psi$ through the trigonometric relationships (see
Fig.~\ref{f3})
\begin{equation}\label{cosi}
   \cos i = \cos i_J \cos\theta_J - \sin\theta_J \sin i_J \cos \Phi \;,
\end{equation}
and
\begin{eqnarray}\label{omega}
  \sin i\sin\omega &=& (\sin\theta_J\cos i_J 
   +\cos\theta_J\sin i_J\cos\Phi)\sin\Psi+\sin i_J\sin\Phi\cos\Psi \;,\\
  \sin i \cos\omega &=& (\sin\theta_J\cos i_J 
   +\cos\theta_J\sin i_J\cos\Phi)\cos\Psi-\sin i_J\sin\Phi\sin\Psi \;. 
\end{eqnarray}
Consequently, due to the non-linear character of these relationships, the
linear-in-time precession of $\Phi$ and $\Psi$ will cause a non-linear-in-time
precessional evolution of the observed parameters $x$ and $\omega$.  Although
the changes in $\omega$ and $x$ can be expressed in a closed analytic form, a
Taylor expansion in powers of time $t-T_0$, where $T_0$ is the time of
periastron passage, is more suitable for observational purposes, as will
become clear in the next section. Thus, one has:
\begin{eqnarray}
 x^{\rm prec}(t) &=& x_0+\dot {x}^{\rm prec}(t-t_0)+
 \frac{1}{2}\ddot{x}^{\rm prec}(t-t_0)^2+\dots\;, \\
   \omega^{\rm prec}(t) &=& \omega_0+\dot{\omega}^{\rm prec}(t-t_0)
                         +\frac{1}{2}\ddot{\omega}^{\rm prec}(t-t_0)^2+\dots\;,
\end{eqnarray}
where $x_0$, $\omega_0$ are initial values of $x$, $\omega$ at the initial
epoch $T_0$,
\begin{eqnarray}
   \dot {x}^{\rm prec}&=&\dot x_S+\dot x_Q\; \\
   \dot{\omega}^{\rm prec}&=&\dot\omega_{GR0}+\dot\omega_S+\dot\omega_Q\;,
\end{eqnarray}
and subindeces $S$ and $Q$ mean that the corresponding quantity is caused by
the influence of the BH spin or the BH quadrupole moment, respectively. The
quantity $\dot\omega_{GR0}$ describes the standard relativistic periastron
advance.

As mentioned previously, the contribution of the quadrupole moment of the BH
companion to any observable parameter is extremely small. For this reason, we
again omit, in what follows, all quantities with the subindex $Q$.  Further
simplification is achieved if one makes use of the fact that, in general, $
S\ll L$ and, thus, as a consequence of equation (\ref{theta}),
$\theta_J\approx S\sin\theta/L \ll 1$.  Indeed, when ${\cal M}_\bullet \ga
10$,
\begin{equation}
   \frac{S}{L} \approx 0.01\;\frac{\chi}{(1-e^2)^{1/2}}\;
               \left(\frac{P_b}{1\;{\rm day}}\right)^{-1/3}
               \left(\frac{{\cal M}_\bullet}{10}\right)^{4/3} \;.
\end{equation}
Therefore, even for a binary system with an orbital period $P_b$ of 0.5 days,
an eccentricity $e$ of 0.8, and a 20$M_\odot$ BH companion $\theta_J$ will be
smaller than $\sim 3\arcdeg$.  Taking this into account equations (\ref{cosi})
and (\ref{omega}) assume the form:
\begin{equation}
   \sin i = \sin i_J+\theta_J\cos i_J\cos\Phi + {\cal O}(\theta_J^2)
\end{equation}
and
\begin{equation}
   \omega = \Psi+\Phi-\theta_J\cot i_J\sin\Phi+{\cal O}(\theta_J^2)\;,
\end{equation}
(cf.\ \cite{Wex98}). Thus, one obtains, to leading order in $\theta_J$,
\begin{eqnarray}
   \dot\omega_S        &\simeq& -B^\star_S\;\chi\sin\theta\cos\Phi_0
      -2A_S^\star\;\chi\cos\theta \;, \label{omdotso}\\
   \frac{\dot x_S}{x}  &\simeq& -B^\star_S\; \chi\sin\theta\sin\Phi_0\;,
      \label{xdotso}\\
   \ddot\omega_S       &\simeq& + C_S^\star\;\chi\sin\theta\sin\Phi_0\;,
      \label{omdotdotso}\\
   \frac{\ddot x_S}{x} &\simeq& -C_S^\star\;\chi\sin\theta\cos\Phi_0\;,
      \label{xdotdotso}
\end{eqnarray}
where $\Phi_0$ is the numerical value of the angle $\Phi$ at the initial epoch
$T_0$ and the constant coefficients
\begin{equation}
   A_S^\star \equiv \Omega_S^\star/\chi\;,\qquad
   B_S^\star \equiv A_S^\star\cot i    \;,\qquad
   C_S^\star \equiv B_S^\star\Omega_S^\star L/S
\end{equation}
are independent of the angles $\theta$, $\Phi_0$, and the spin parameter
$\chi$. The values for these coefficients are fixed once we know the masses of
the binary system.

At the beginning of the timing project one will just be able to see the linear
trend in the precession, $\dot\omega$, and some time later $\dot x$. But
after a while, depending on the compactness of the binary system, one will
start to see $\ddot\omega$ and $\ddot x$. Since $\dot\omega$ is clearly
dominated by $\dot\omega_{GR0}$ the first sign for the existence of spin-orbit
coupling in the binary system will come from the measurement of $\dot
x$. Under favorable circumstances (see equation (\ref{xdot3})
and related discussion) this allows us to calculate $\chi\sin\theta\sin\Phi_0$ if we assume
that the masses of pulsar and BH and, thus, the angle of the orbital
inclination, $i$, are known from the measurement of at least two PK
parameters.  Since $|\chi\sin\theta\sin\Phi_0| \le |\chi|$ we can get
a lower limit for the spin of the BH or, if we find $|\chi|>1$, a Kerr BH is
ruled out. In fact, it is interesting to note that if one assumes that
the precession of PSR J0045--7319 (see Kaspi {\it et al.}\ 1996) is caused by
a BH companion, the analysis presented here would give $|\chi|\ga100$ which is
certainly not a Kerr BH.\footnote{There is clear observational evidence that
the companion of PSR J0045--7319 is a B star (Bell {\it et al.}\ 1995). The
precession of the orbit is caused by the large quadrupole moment of the B star
(Kaspi {\it et al.}\ 1996)} The measurement of $\ddot x$ will then allow the
separate determination of the angle $\Phi_0$ and $\chi\sin\theta$. If we can
determine the masses of the bodies comprising the binary system without
measuring $\dot\omega$ (e.g., using $\gamma$ and $\dot P_b$) then, in
principle, we can extract $\dot\omega_S$ from the total observed value of
$\dot\omega$ and finally determine $\chi$. It is important to see that
neither $\ddot\omega$ nor even higher derivatives of $x$ and $\omega$ contain
a term $\chi\cos\theta$. Thus the only way to fully determine the spin of the
BH is the separate measurement of $\dot\omega_S$.

Unfortunately, the precession of the orbit due to spin-orbit coupling is not
the only effect that gives rise to the parameter $\dot x$ in the timing model.
Following Damour \& Taylor (1992) and Kopeikin (1994, 1996) one finds, in
fact, four alternative effects causing a secular change of the observed value,
$x^{\rm obs}$, of the parameter $x$:
\begin{equation}\label{xdot3}
   \left(\frac{\dot x}{x}\right)^{{\rm obs}} =
      \left(\frac{\dot x}{x}\right)^{{\rm prec}}
      +\left(\frac{\dot a_p}{a_p}\right)^{{\rm gw}}
      +\left(\frac{\dot x}{x}\right)^{{\rm pm}}
      +\frac{d\varepsilon_A}{dt}-\frac{\dot D}{D} \;.
\end{equation} 
The first contribution to the right-hand side of equation (\ref{xdot3}) is due
to the precession of the binary orbit discussed at the beginning of this
section. The contribution to the right-hand side of equation (\ref{xdot3})
from the second ({\rm gw}) term is caused by the shrinking of the pulsar orbit
due to the emission of gravitational waves by the binary system. Given the
masses of pulsar and companion this effect can be calculated (\cite{Pet64})
and subtracted. The third (pm) contribution is due to the proper motion of the
binary system, which causes a gradual change in the apparent geometrical
orientation of the orbital plane (Kopeikin 1994, 1996; see also
\cite{AJRT96}).\footnote{Apart from a contribution to $\dot x$ this effect
changes also the observed value for the advance of periastron, $\dot\omega$
(Kopeikin 1994, 1996).}  If the pulsar is close to the solar system and/or has
a high proper motion this contribution can become quite significant, in which
case, additional astrometric information will be needed for measurement and
subtraction of $\dot{x}^{\rm pm}$.  The forth contribution to the right-hand
side of the equation (\ref{xdot3}) is due to a varying aberration caused by
the deSitter precession of the pulsar's spin while the pulsar orbits its
companion.  The pulsar spin precesses around the direction of the orbital
angular momentum, ${\bf L}$, with angular velocity (\cite{BOC75})
\begin{equation}\label{deSp}
   \Omega^{{\rm spin}}_p=
      \frac{\tsun^{2/3}}{1-e^2}\;\left(\frac{2\pi}{P_b}\right)^{5/3}
      \frac{\Mb(4\Mp+3\Mb)}{2(\Mp+\Mb)^{4/3}}\;
\end{equation}
and, consequently, (\cite{DT92})
\begin{equation}
   \frac{d\varepsilon_A}{dt} = 
      -\frac{P}{P_b}\;\frac{\Omega^{{\rm spin}}_p}{(1-e^2)^{1/2}}\;
      \frac{\cot\lambda_p\sin2\eta_p+\cot i\cos\eta_p}{\sin\lambda_p} \;,
\end{equation}
where $P$ is the pulsar's rotational period. A comparison with equation
(\ref{xdotso}) leads to (excluding values of $\lambda_p$ close to 0\arcdeg and
180\arcdeg)
\begin{equation}
   \left|\frac{\dot\varepsilon_A}{\dot x_S/x}\right| \sim 
      \frac{P}{P_b}\frac{\Omega^{{\rm spin}}_p}{\Omega^\star_S} \sim 0.006
      \left(\frac{P}{1\;{\rm s}}\right)
      \left(\frac{P_b}{1\;{\rm day}}\right)^{-2/3}
      \left(\frac{\Mb}{10}\right)^{-1/3} \;.
\end{equation}
Thus, the varying aberration effect will be of the same order as the
precession of the orbit only for slow pulsars ($P\ga1$ s) in very close orbits
with a BH companion.  But in this case the varying aberration will also show
up as a change of the eccentricity (\cite{DT92})
\begin{equation}
   \left(\frac{\dot e}{e}\right)^{{\rm obs}} = 
      \left(\frac{\dot e}{e}\right)^{{\rm gw}}+\frac{d\varepsilon_A}{dt}\;,
\end{equation}
and can therefore be observed and subtracted. (The change due to gravitational
wave damping (gw) can be calculated if the masses are known (see
\cite{Pet64}).)

The last term in equation (\ref{xdot3}) is due to a varying Doppler shift,
$D$, caused by a secular change of the distance between the solar-system
barycenter and the binary-pulsar system.  The effect under discussion is
non-linear in time and can be caused both by the acceleration of gravitational
fields in our galaxy and an apparent acceleration due to the proper motion of
the binary system in the sky (\cite{Shk70}; \cite{DT91}):
\begin{equation}
  -\frac{\dot D}{D}=\frac{1}{c}\;{\bf K}_0\cdot({\bf a}_{{\rm binary\;system}}
  -{\bf a}_{{\rm solar\;system}})+\frac{V_T^2}{cd}\;.
\end{equation}
Here, $V_T$ denotes the velocity of the pulsar being transverse to the line of
sight, $d$ is the distance between the pulsar and the solar system, and ${\bf
a}_{{\rm binary\;system}}$ and ${\bf a}_{{\rm solar\;system}}$ are galactic
accelerations at the locations of the solar and binary systems
respectively. The galactic acceleration at the solar location is of order
$2\times10^{-8}\;{\rm cm}\;{\rm s}^{-2}$ which is a typical value for the
acceleration fields in the galactic disk (\cite{CI87}; \cite{KG89} for a model
of the Galactic acceleration field). Hence, unless the pulsar is located in
the core of a globular cluster where accelerations are much higher, the
contribution of galactic gravitational fields to $\dot x$ is negligible. The
same is true for the Shklovskii term if we assume typical distances ($\ga 1$
kpc) and typical pulsar velocities (few 100 km/s). Thus, we conclude that a
varying Doppler shift should not complicate the interpretation of timing
observations of BH-pulsar binaries under ordinary circumstances.

As pointed out above, for a full determination of the rotational parameter
$\chi$ of the pulsar companion, one cannot make use of $\dot\omega$ to
calculate the masses of pulsar and BH. Thus, one needs a third PK
parameter for the separate mass determination which will be most likely $\dot
P_b$ (caused by gravitational wave damping). Since (\cite{DT91})
\begin{equation}
   \left(\frac{\dot P_b}{P_b}\right)^{{\rm obs}} = \left(\frac{\dot
   P_b}{P_b}\right)^{{\rm gw}}-\frac{\dot D}{D}\;,
\end{equation}
and
\begin{equation}
    \left(\frac{\dot P_b}{P_b}\right)^{{\rm gw}} \approx
    \left(-1.6\times10^{-18}\;{\rm s}^{-1}\right)\; \left(\frac{P_b}{1\;{\rm
    day}}\right)^{2/3}\; \left(\frac{\Mb}{10}\right)^{2/3}\;
\end{equation}
one has to be cautious about the contribution of the galactic acceleration and
Shklovskii's term.  Even if we understand these effects to about 10\% for
BH-pulsar binaries with orbital periods greater than a few days, a reliable
determination of $\dot{\omega}_S$ will be impossible. Exceptions are cases
where the system is seen sufficiently edge-on so that one can measure the two
PK parameters related with the Shapiro delay, $r$ and $s$, with necessary
precision. For PSR B1913+16, the limit on the determination of $\dot P_b^{{\rm
gw}}$ due to galactic accelerations has been clearly demonstrated by
Damour \& Taylor (1991).

For a BH-pulsar binary the deSitter precession of the pulsar spin, equation
(\ref{deSp}), is approximately $\dot{\omega}_{PN}/2$. For short orbital
periods this can be a few degrees per year or even more. Then, if the pulsar
spin is tilted with respect to the orbital angular momentum the deSitter
precession can lead to significant changes in the pulse profile (cf.\
\cite{WRT89}; \cite{Kra98}) which might, in principle, limit the timing 
accuracy. In extreme cases the pulsar could even cease to beam in our
direction after a couple of years.

Only if the influence of all these additional effects listed above is well
understood we can hope to perform a spin determination for the BH companion of
the pulsar.


\section{Parameter estimations}

So far we have shown that in principle, one can use the pulsar timing
observations of a BH-pulsar binary system to determine the mass and spin of
the rotating BH companion. In the previous section, we discussed in detail the
extraction of the BH mass and spin from the observed values of Keplerian and
PK parameters. However, the number of measured parameters one hopes to
determine from timing observations depends critically on both the accuracy of
the measured TOAs and the compactness of the binary system's orbit.  For
example, we show that an ordinary one-second period pulsar in a one-year orbit
with a 10 $M_\odot$ BH will allow at most the determination of the mass
function and sets only a lower limit on the mass of the BH companion. However,
for a millisecond pulsar in an eccentric one-day orbit with a $10M_\odot$ BH,
the masses of the pulsar and companion can be determined with high precision
after a month of regular timing observations from the measurement of two PK
parameters (most likely $\dot\omega$ and $\gamma$). After a one-year observing
campaign one can expect to see additional relativistic effects, such as
gravitational radiation damping and relativistic spin-orbit precession. In
this section, we explore the potential measurability of various relativistic
effects in BH-pulsar binaries.  We use a standard tool, the information
matrix, to asses the errors of our parameter estimation method.

Let $N$ denote the total number of measured TOAs, $\tau_i$ ($i=1,2,...,N$),
and $M$ the number of fitting parameters, $\xi_a$ ($a=1,2,...,M$), in the
timing model used for the a priori estimate of ${\cal N}(\tau,\bm{\xi})$ the
number of the pulse arriving at the time $\tau$.  The parameters are
determined using a least-square minimization method where the goodness of fit
parameter, $\chi^2$, is given by
\begin{equation}
   \chi^2 = \sum_{i=1}^N\left[\frac{n_i-{\cal N}(\tau_i,\bm{\xi})}
            {\nu_p\;\sigma_i}\right]^2 \;.
\end{equation}
Here $\nu_p$ is the pulsar's rotational frequency, $n_i$ is the closest
integer to ${\cal N}(\tau_i,\bm{\xi})$, and $\sigma_i$ is the estimated
uncertainty of the $i$-th TOA.  The information matrix is defined as
\begin{equation}
   L_{ab} = \sum_{i=1}^{N}\frac{\partial{\cal N}(\tau_i)}{\partial\xi_a}\;
                          \frac{\partial{\cal N}(\tau_i)}{\partial\xi_b}\;.
\end{equation}
If only white noise is present in the TOA residuals, the inverse of the
information matrix, $M_{ab} \equiv L_{ab}^{-1}$, is the correlation matrix of
the fitting parameters. The elements of the main diagonal of $M_{ab}$ give the
variations of the measured parameters, $\langle\xi_a^2\rangle=M_{aa}$, and the
off-diagonal elements of $M_{ab}$ represent the correlations between them
(\cite{Bar74}).

Two kinds of BH-pulsar binaries seem likely to be present in our Galaxy.
First, young binary pulsars located in the Galactic plane, where the pulsar is
the result of the second supernova explosion. Numerical simulations indicate,
that such a pulsar is expected to be in eccentric wide ($P_b\ga 10$ days)
orbits (\cite{LPPO94}). Secondly, a millisecond pulsar captured by a BH, most
likely located in a globular cluster (\cite{KHMcM93}).\footnote{The ratio of
orbital binding energy to potential energy in the gravitational field of the
cluster is only a very weak function of the companion mass, once
$m_c\ga1.4M_\odot$. Thus, BH-pulsar binaries should have roughly the same
probability to remain in or near the cluster after formation as NS-NS
binaries, like PSR B2127+11C.}  The aim of our numerical estimates is to
understand which parameters one can expect to measure in these two different
types of BH-pulsar binaries within a reasonable time span of observations
($\la 20$ years).  Here we assume that all measured TOAs, $\tau_i$, are of a
similar quality, i.e.\ all $\sigma_i$ are the same ($\sigma_i=\sigma_{{\rm
TOA}}$). Furthermore, there is at least one timing observation every month
after an initial set of 10 timing observations along one full orbit which give
the Keplerian parameters required for subsequent exploration of the binary
system. We further assume that the pulsar is in a highly eccentric orbit
($e=0.8$) with a $10M_\odot$ BH companion.

Figure~4 shows the estimated time span of regular timing observations,
$T_{{\rm obs}}$, needed for the measurement of certain astrophysical parameters
or effects, as a function of the orbital period, $P_b$. As astrophysical
parameters and effects we consider:
\begin{itemize}
\item Total mass of the binary system --- measurement of $\dot\omega$.
\item Mass of the BH --- measurement of $\dot\omega$ and $\gamma$.
\item Testing the emission of gravitational waves --- measurement of 
      $\dot\omega$, $\gamma$ and $\dot P_b$ (cf.\ \cite{TW89}).
\item Frame-dragging effect --- measurement of $\dot x$.
\item Spin of the BH (magnitude and orientation) --- measurement of 
      $\dot\omega$, $\gamma$, $\dot P_b$, $\dot x$, and $\ddot x$. The mass
      determination has to be done using $\gamma$ and $\dot P_b$.  The
      accuracy for the total mass has to better than
      $\sim\varepsilon\dot\omega_S/\dot\omega$, where $\varepsilon$ denotes
      the aspired accuracy for the BH spin measurement.
\end{itemize}
In figure~4a we have assumed a timing accuracy of $\sigma_{{\rm TOA}} \sim
100\;\mu$s which represents the optimum one can expect for timing
observations of a young pulsar, with a typical rotational period of 100 ms to
a few seconds. In figure~4b we have assumed the presence of a millisecond
pulsar, i.e.\ $\sigma_{{\rm TOA}} \sim 1\;\mu$s.

We conclude from figure~4a that for a young pulsar only in systems with
orbital periods below 0.2 days the measurement of the spin of the BH
(magnitude and orientation) is possible within a reasonable observing
time span. For systems with orbital periods below 3 days one can
expect to see the frame-dragging caused by the extreme Kerr BH, i.e.\ measure
$\dot x$. Systems with orbital periods below 10 days will allow a mass
determination for the BH with an accuracy of better than 5\% after less than
20 years of observing. The periastron advance will be a prominent effect even
for orbital periods exceeding one year and thus the total mass of the binary
system should be a measureable quantity for all orbital periods $\la$1000
days.

In the case of a millisecond pulsar in orbit with a 10 solar mass extreme Kerr
BH, where one expects timing accuracies two orders of magnitude better than in
the previous case, the corresponding time spans for observations are clearly
shorter (see figure~4b). Therefore the spin of the BH will be measurable for
systems with orbital periods up to one day, and the frame-dragging will be
seen even in systems with 10 days orbital period. For systems with orbital
periods below a hundred days a good mass estimates for the BH companion will
be possible, and the total mass of the binary system will be measurable even
in very wide systems with orbital periods of a few years.

The previous calculations were done for a BH mass of $10M_\odot$, which we
assume to be typical for stellar-mass BHs. In a $30M_\odot$ BH-pulsar binary
spin-related effects are expected to be an order of magnitude larger, since
the spin of a BH scales with its mass squared (see
equation~(\ref{spinbh})). On the other hand, there might be a fairly large
population of BHs in our Galaxy with masses slightly above $1.5M_\odot$
(\cite{BB94}). Our calculations show that only for very short orbital periods
($P_b\la0.2$ days) and high orbital eccentricities, ($e\ga0.8$) one can expect
to see the relativistic spin-orbit precession ($\dot x_S$) within less than 20
years of observing, given an uncertainty in the TOAs which is not worse than
$1\mu$s. In this case the measurement of $\dot x_S$, which gives a lower limit
for $\chi$, might have the potential to distinguish between a NS and a BH,
since it seems that $\chi\la0.7$ for NSs of comparable mass (\cite{LP97}). It
would require that the angle $\theta$ is close to $90\arcdeg$ (the spin of the
BH lies almost in the orbital plane) and $\Phi_0$ is either close to
$90\arcdeg$ or close to $270\arcdeg$. Then, if $\chi$ is close to one for the
BH companion, the minimum value for $\chi$ derived from the $\dot x$
measurement can exceed the maximum value for NSs.

Let us point out that on time scales of several years, timing observations are
usually contaminated by the increasing amount of red (low frequency) timing
noise (\cite{CH80}, \cite{Tay91}, \cite{Lyn96}, \cite{Kop97b}). If the red
noise dominates in TOAs residuals, it makes the problem of measuring
relativistic parameters more difficult. In particular for young pulsars in
wide orbits, the separation of orbital effects and timing noise may cause a
severe problem (cf.\ \cite{Wex&al98}).


\section{Measuring the quadrupole of the BH companion}

Equation (\ref{nohair}) implies that with the determination of the mass and
the spin of a Kerr BH we also know its quadrupole moment. Therefore, if we are
able to extract independently the quadrupole moment of the companion from our
timing observations we could actually test whether the observed pulsar is
orbiting a Kerr BH or another compact relativistic object like neutron or
boson star.  As discussed in Section 4 the quadrupole moment of the BH
companion will lead to an additional precession in the angles $\Phi$ and
$\Psi$ and, thus, to a secular time evolution in the parameters $\omega$ and
$x$.  Unfortunately these secular changes in the orientation of the orbit
caused by the quadrupole moment of a BH companion are typically three orders
of magnitude smaller than the changes caused by the relativistic spin-orbit
coupling (see equation (\ref{OMqn})).  Thus, the chance of extracting the
quadrupole moment of the BH from the measurement of the parameters
$\dot\omega$ and $\dot x$ is unlikely. On the other hand, the anisotropic
nature of the quadrupole component of the external gravitational field
(\ref{asympt1}) will lead to characteristic short-term periodic effects every
time the pulsar gets close to the oblate companion.  A detailed investigation
of these short-term effects can be found in Wex (1998). These short-term
periodic effects lead to a unique signature, which will show up in the
post-fit TOA residuals if one uses a timing model that accounts only for the
secular changes of the precession and the short-term periodic effects caused
by the mass-monopole (\cite{DD86}) and spin (\cite{Wex95}) of the BH.
Figure~5a shows the result of numerical simulations of TOA residuals for
a pulsar orbiting a 30 $M_\odot$ BH in an 0.1 day orbit with an eccentricity
of 0.9. This represents certainly an unrealistic case from the observational
point of view since the lifetime of such a system due to gravitational
radiation damping is just $10^4$ years which makes a discovery of such a
system extremely unlikely.  In addition, the accuracy needed to measure this
effect would require a nano-second timing precision.  Even for the best
millisecond pulsars the present timing precision is still more than two orders
of magnitude worse and, thus, we conclude that the quadrupole of a
stellar-mass BH companion is not measurable in timing observations. However,
since the quadrupole moment of a BH scales with the mass cubed (see equation
\ref{quadrubh}), these short-term periodic effects under discussion gain
importance for very massive BHs which could sit in the center of globular
clusters as a result of a collapse of the cluster core.  Figure~5b gives
numerical simulations of TOA residuals for a pulsar orbiting a $10^4$
$M_\odot$ BH in a 10 day orbit with an eccentricity of 0.9. A timing accuracy
of a few hundred nanoseconds would allow the measurement of the quadrupole
moment. Of course, even higher BH masses and/or smaller pulsar orbits will
make the quadrupole a prominent feature in the TOAs.


\section{Summary}

In this paper we have examined how timing observations of a pulsar orbiting a
rotating stellar mass BH can be used to study the physical properties of the
BH companion. We pointed out that the measurement of two post-Keplerian
parameters can lead to a mass determination of the black hole with very high
accuracy, far better than it is possible for present black hole candidates. It
was shown that the frame dragging propagation effect discussed by Laguna \&
Wolszczan (1997) will not be separable from the bending delay for those BH
companions less than a hundred solar masses, since it would require
determining the orientation of the pulsar's spin with an accuracy that is not
likely possible with pulse structure analysis. We gave detailed calculations
about the (secular) influence of the BH rotation on the orbital motion of the
pulsar.  In particular the long term behavior of this relativistic spin-orbit
coupling can be described by two angles, defined with respect to the
invariable plane, which change linearly in time. For an observer this
linear-in-time precession converts into a non-linear in time evolution of the
orbital inclination and the longitude of periastron. In a timing model for
such a binary system, this can be taken into account by fitting for the
parameters $\dot x$, $\ddot x$, $\dot\omega$, and $\ddot\omega$ as is done for
pulsars with main-sequence star companions where the classical spin-orbit
coupling is the cause of the precession (see \cite{Wex98}). The measurement of
$\dot x$, $\ddot x$, and $\dot\omega$ could, in principle, lead to the
determination of the BH spin, if the masses of the BH and the pulsar can be
determined without making use of the advance of periastron, using, for
example, the measurement of the Einstein delay ($\gamma$) and the orbital
period decay ($\dot P_b$). In order to do so, a fractional measurement
precision of $\la 10^{-4}$ is required for $\dot\omega$, $\gamma$, and $\dot
P_b$. If this measurement precision is not achieved, or if there are Galactic
and kinematic effects present which are not sufficiently understood, then only
the projection of the BH spin onto the orbital angular momentum will be
measurable. If the timing data allow just the determination of $\dot x$ then
we can calculate only a lower limit for the projected BH spin, which, of
course, is also a lower limit for the BH spin. However if we have just the
lower limit for the spin, the presence of the $\dot x$ would already indicate
frame dragging if an oblate non-compact companion can be ruled out (for
example by optical observations, the absence of eclipses of the pulsar,
absence of tidal effects, etc.).  On the other hand, the method under
discussion can be used to rule out a Kerr BH companion as has been
demonstrated here in the case of PSR J0045--7319.

We have studied the measurability of the relativistic spin orbit coupling for
two different cases. First, a young pulsar orbiting a $10 M_\odot$ extreme
Kerr BH and second, a millisecond pulsar orbiting a $10 M_\odot$ extreme Kerr
BH.  In the first case only for binary pulsars with orbital periods less than
three days one can expect to measure $\dot x$, i.e.\ the frame dragging caused
by the rotating BH, with sufficient accuracy, but in the second case a
measurement of $\dot x$ seems very likely. For BH-millisecond pulsar binaries
with an orbital period below one day one can expect a measurement of $\ddot x$
(and $\ddot\omega$) within a reasonable time span of observations, and one
will be able to fully determine the BH spin parameters (magnitude and
orientation). However, if the pulsar is located close to the core of a
globular cluster one might not have the necessary understanding for the
gravitational cluster potential in order to take properly into account its
contribution to the measured binary parameters. For a $1.5M_\odot$
BH-millisecond pulsar binary with $P_b \la 0.2$ days and $e\ga 0.8$ one might
be able to measure $\dot x_S$. In principle, this could help to exclude a
neutron star, if the magnitude and orientation of the BH companion is such
that $\chi\sin\theta\sin\Phi_0\ga0.7$, since NSs seem to have $\chi\la0.7$.

Finally we pointed out that the quadrupole field of the rotating BH will lead
to a distinctive signature in the post-fit TOA residuals during each
periastron passage. However, our numerical simulations suggest that this would
be observable only for very massive BHs, typically more than $10^4 M_\odot$.
Whether BHs of this size do exist, e.g.\ in the center of globular clusters,
is still unclear.  The discovery of a pulsar orbiting the super-massive BH in
the center of our Galaxy could provide a possible setting for these
measurements.


\acknowledgments

We thank B. Paczynski for attracting our attention to the problem, P.  Laguna,
D. Lorimer, S. Thorsett, and A. Wolszczan for valuable discussions, and
R. Dewey for carefully reading the manuscript. Norbert Wex acknowledges the
hospitality of the Department of Physics at Princeton University during tenure
of the Otto-Hahn Prize. S. M. Kopeikin is pleasured to acknowledge the
hospitality of G. Neugebauer and G. Sch\"afer and other members of the
Institute for Theoretical Physics of the Friedrich Schiller University of
Jena.  This work has been partially supported by the Th\"uringer Ministerium
f\"ur Wissenschaft, Forschung und Kultur grant No B501-96060.


\newpage 

\newpage

\newpage

\newpage

\begin{figure}
\centerline{\psfig{figure=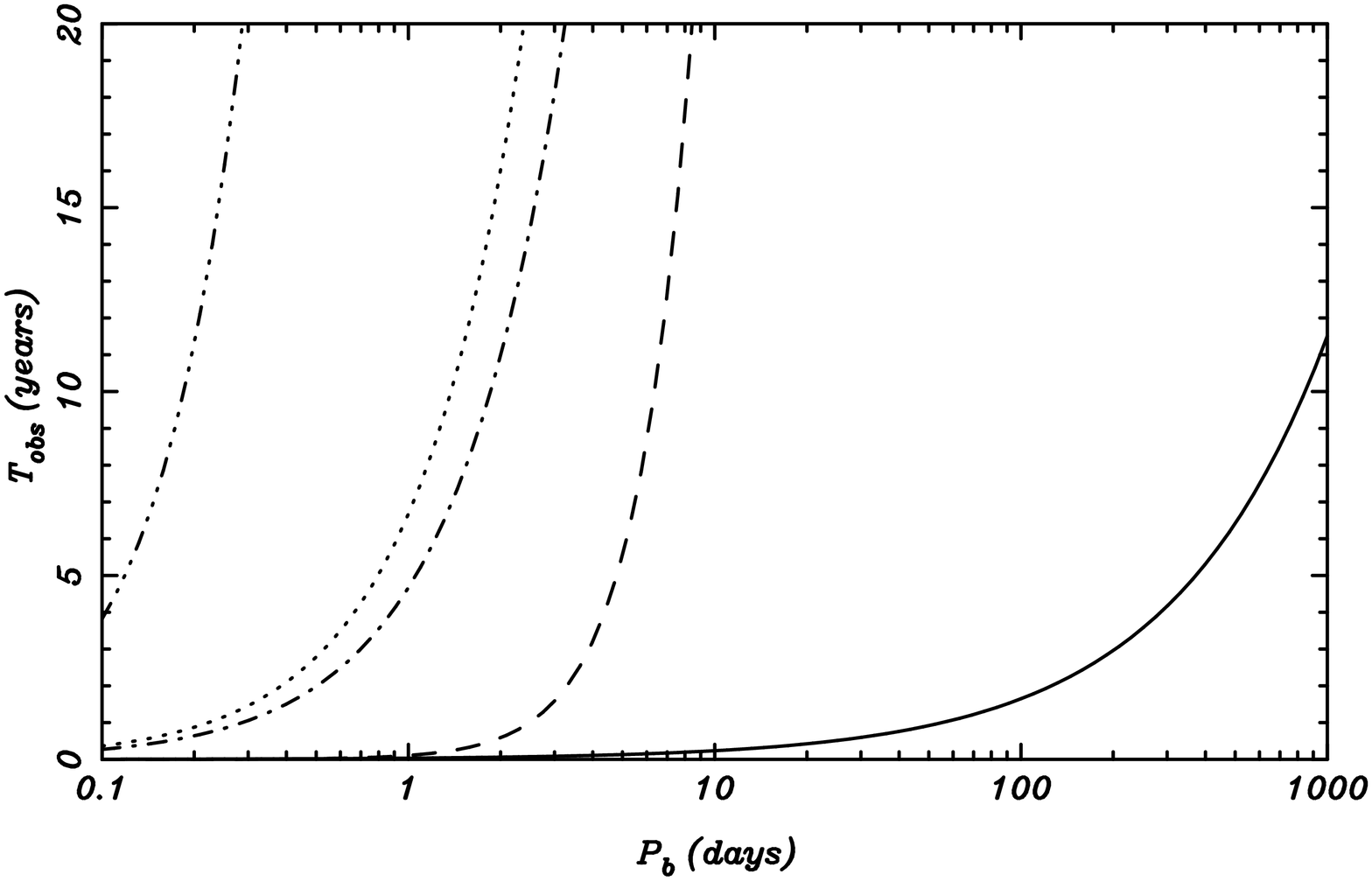,angle=-90,width=5in}}
\centerline{\psfig{figure=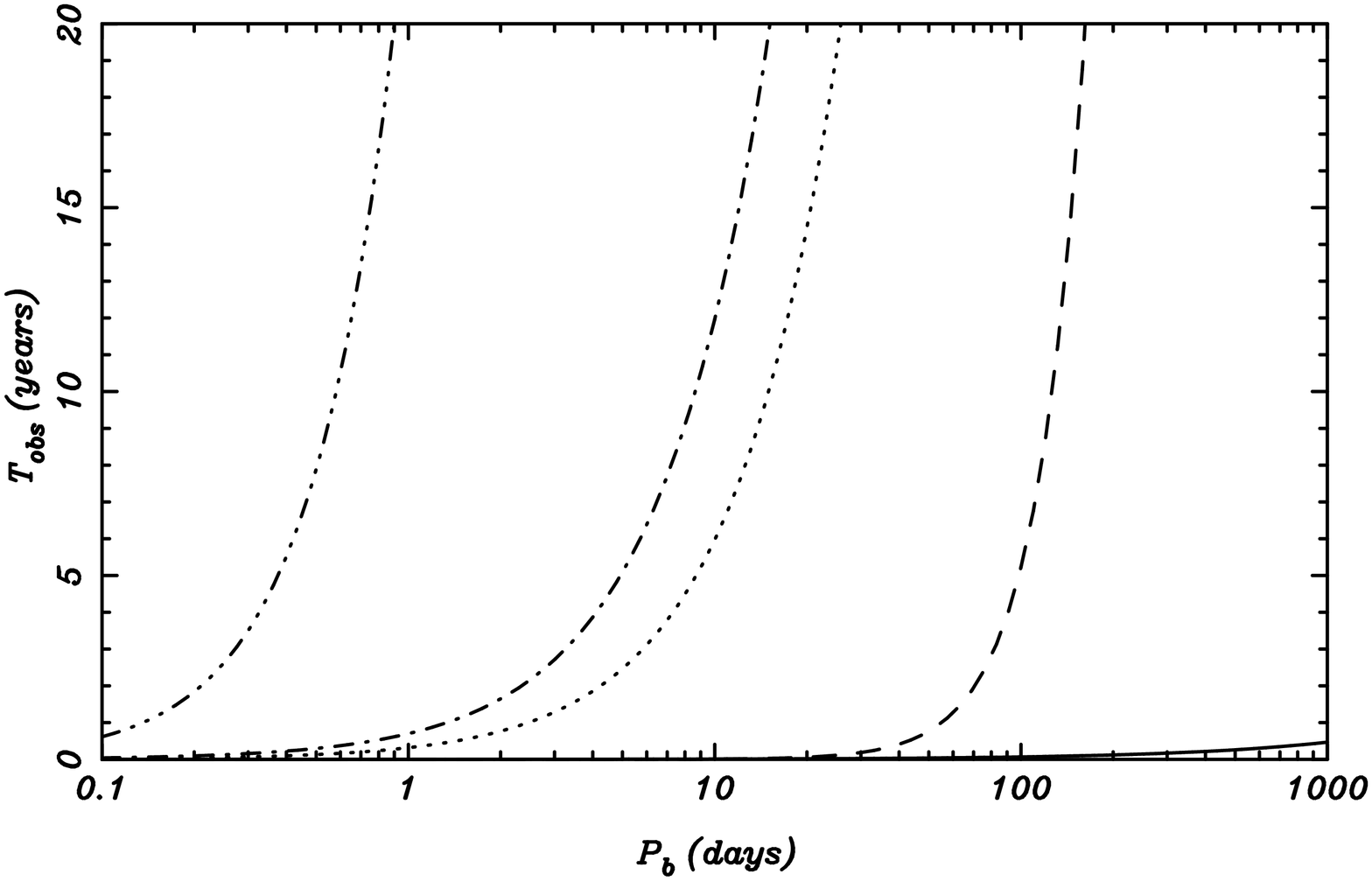,angle=-90,width=5in}}
\caption{
Time span of observations needed (after 10 timing observations along a full
orbit) to determine the total mass of the BH-PSR system with 5\% accuracy
(full), the mass of the BH with 5\% accuracy (dashed), the emission of
gravitational waves with 1\% accuracy (dot-dash), the precession caused by
frame dragging with 5\% accuracy (dotted), and the spin of the BH with 5\%
accuracy (dash-dot-dot-dot) as a functon of the orbital period
$P_b$. Estimations were done for a 10 solar mass extreme Kerr BH with $\theta=
45\arcdeg$ and $\Phi_0=45\arcdeg$, an orbital eccentricity of 0.8 and a timing
accuracy, $\sigma_{{\rm TOA}}$, of 100 $\mu$s (upper figure) and 1 $\mu$s
(lower figure).  }

\label{f4}
\end{figure}

\newpage

\begin{figure}
\centerline{\psfig{figure=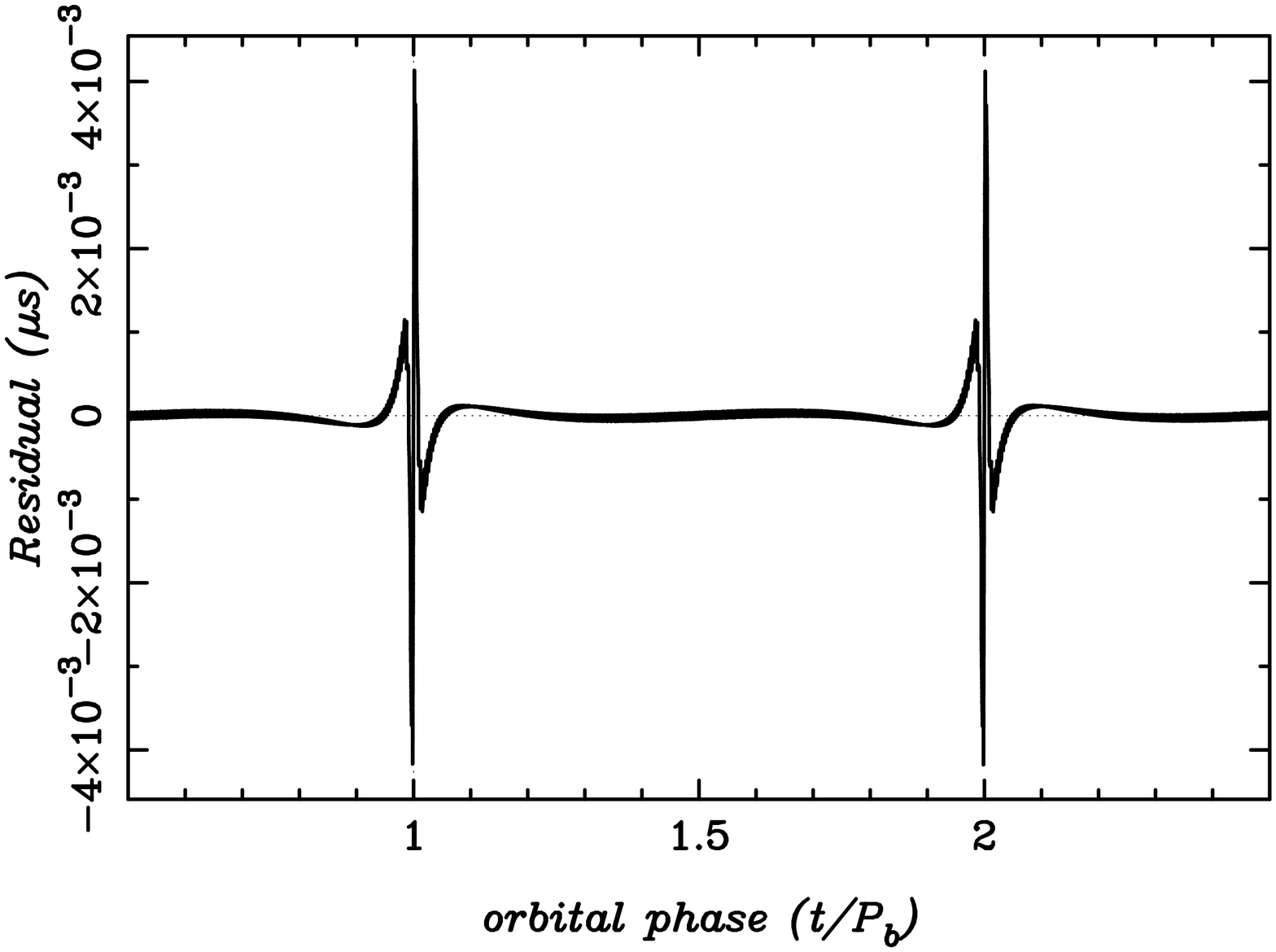,angle=-90,width=5in}}
\centerline{\psfig{figure=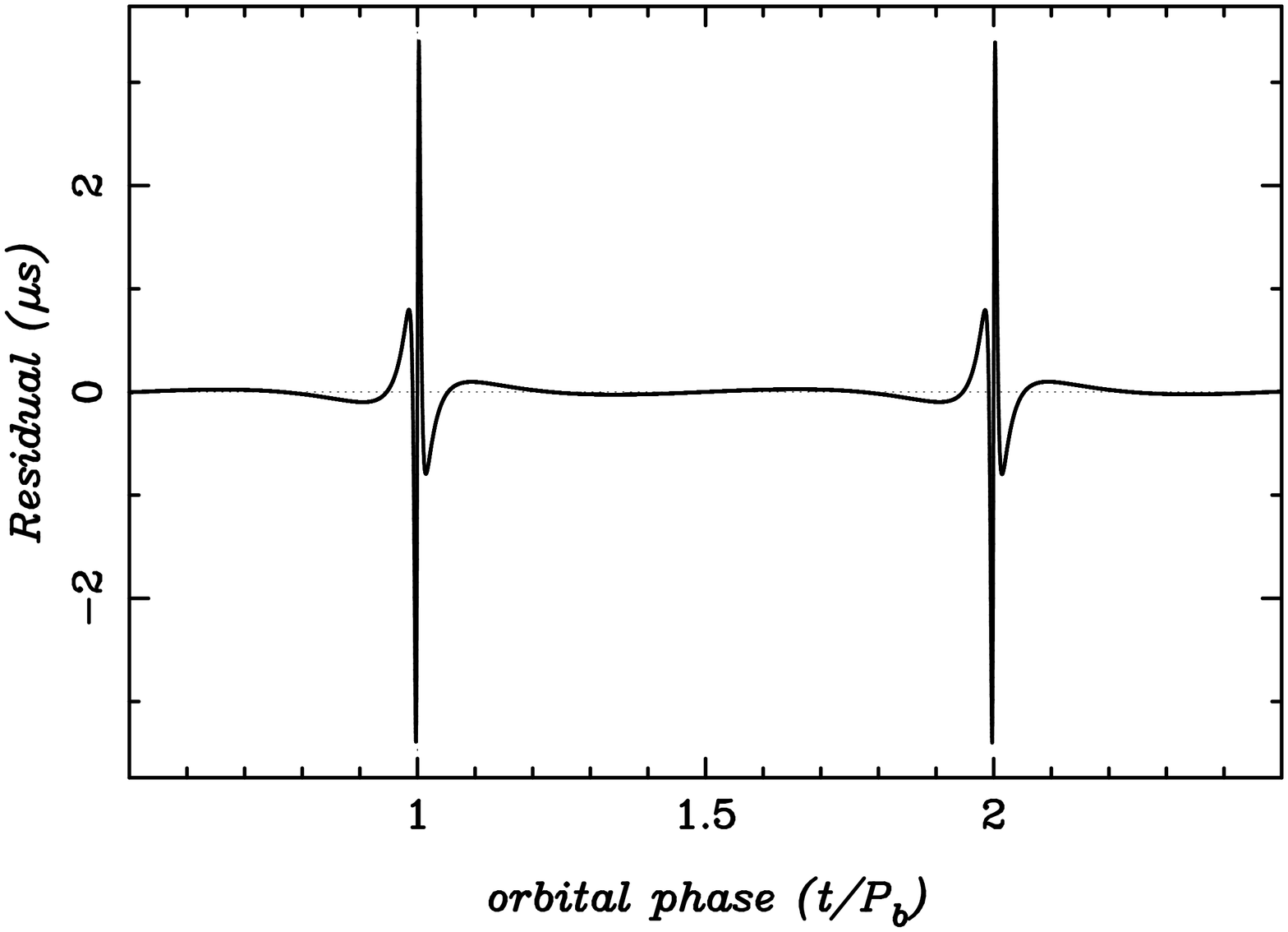,angle=-90,width=5in}}
\caption{
{\it Upper figure:} Typical signature in the timing residuals caused by the
quadrupole moment of a 30 $M_\odot$ BH companion. We used $P_b=0.1$ days and
$e=0.9$.  The inclination of the BH spin with respect to the orbital plane
(the angle $\theta$) was assumed to be $70\arcdeg$. {\it Lower figure:} Timing
residuals caused by the quadrupole moment of a $10^4$ $M_\odot$ BH
companion. We used $P_b=10$ days and $e=0.9$. The inclination of the BH spin
with respect to the orbital plane (the angle $\theta$) was assumed to be
70\arcdeg.  
}
\label{f5}
\end{figure}

\end{document}